\documentclass[10pt,aps,pra,twocolumn,superscriptaddress,showpacs]{revtex4-1}
\usepackage{amsmath}
\usepackage{amsfonts}
\usepackage{graphicx} % Include figure files
\usepackage{float}
\usepackage{dcolumn}
\usepackage{bm}
\usepackage{color}

\begin{document}

\newcommand{\li}[1]{ {\color{red} #1}}

\title{Entanglement and entropy production in coupled single-mode Bose-Einstein condensates}
\author{Izabella Lovas}
\affiliation{MTA-BME Exotic Quantum Phases ``Momentum'' Research Group and Department of Theoretical Physics, Budapest University of Technology and Economics, 1111 Budapest, Hungary}
\author{J\'ozsef Fort\'agh}
\affiliation{CQ Center for Quantum Science, Physikalisches Institut, Eberhard Karls Universit\"at T\"ubingen, Auf der Morgenstelle 14, D-72076 T\"ubingen, Germany}
\author{Eugene Demler}
\affiliation{Physics Department, Harvard University, Cambridge, Massachusetts 02138, USA}
\author{Gergely Zar\'and}
\affiliation{MTA-BME Exotic Quantum Phases ``Momentum'' Research Group and Department of Theoretical Physics, Budapest University of Technology and Economics, 1111 Budapest, Hungary}

\date{\today}

\begin{abstract}
We investigate the time evolution of the entanglement entropy of coupled single-mode Bose-Einstein condensates in a double well potential at $T=0$ temperature, by combining numerical results with analytical approximations. We find that the coherent oscillations of the condensates result in entropy oscillations on the top of a linear entropy generation at short time scales. Due to dephasing, the entropy eventually saturates to a stationary value, in spite of the lack of equilibration. We show that this long time limit of the entropy reflects the semiclassical dynamics of the system, revealing the self-trapping phase transition of the condensates at large interaction strength by a sudden entropy jump. We compare the stationary limit of the entropy to the prediction of a classical microcanonical ensemble, and find surprisingly good agreement in spite of the non-equilibrium state of the system. Our predictions should be experimentally observable on a Bose-Einstein condensate in a double well potential or on a two-component condensate with inter-state coupling.
\end{abstract}


\maketitle

\section{Introduction}

Entanglement is a fundamental concept of quantum mechanics, manifesting in strong, non-local correlations between subsystems. Constituting one of the most crucial differences between classical and quantum physics, entanglement is studied in a diverse area of physics, ranging from quantum gravity \cite{blackhole} to topological order in condensed matter systems \cite{topology}. In recent years, entanglement generation in non-equilibrium many-body systems received a special attention, due to the intimate connection between entanglement spreading and the equilibration in closed systems \cite{NonEquilibriumReview,NonEquilibrium,schmiedmayer,trotzky}. Even for globally pure quantum states, the generation of strong entanglement between subsystems allows the thermalization of an isolated quantum system under its own coherent dynamics in a sense that measurements of local observables become indistinguishable from the predictions of an equilibrium thermal ensemble \cite{thermalization,thermalization2,thermalization3}. In contrast, the large number of conserved quantities in integrable systems can prevent entanglement spreading, and result in the failure of thermalization. The slow, logarithmic increase of entanglement has been suggested as a fingerprint of non-ergodic many-body localized phases \cite{mbllog,mbllog2,mbllog3}, whereas delocalized phases are characterized by a linear, light cone-like propagation of correlations \cite{lightcone,ballistic}. 

Despite its fundamental importance, the experimental investigation of entanglement in correlated many-body systems remains challenging, since it usually requires information on the full quantum state. However, the swift experimental progress in recent years opened up unprecedented possibilities to study entanglement in ultracold atomic settings \cite{ZwergerReview}. Site-resolved control of ultracold atoms in optical lattices allowed the direct measurement of R\'enyi entanglement entropy and mutual information \cite{lightcone2,greiner1,bosonentanglement}, as well as the investigation of the intimate relation between the quantum purity of subsystems and the thermalization of an isolated non-equilibrium system \cite{greiner2}.

\begin{figure}[b!]
\includegraphics[width=0.5\columnwidth]{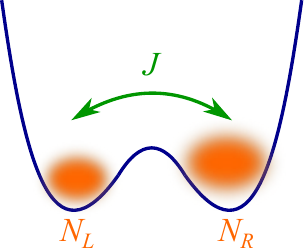}\\
\caption{Illustration of the tunnel coupling described by Hamiltonian \eqref{eq:Hboson}. A Bose-Einstein condensate is loaded into a double well potential, with tunneling $J$ between the two sides. The $N_L$ and $N_R$ particles on the left and right hand side condense into a single wave function. Bosons in the same well repel each other with interaction strength $U$.}
\label{fig:intro0}
\end{figure}

Entanglement is also at the heart of entropy production in closed quantum systems \cite{deutsch,rigol}. Taking two coupled quantum systems, even coherent evolution of the whole system produces entanglement entropy for each subsystem and may lead to equilibration. Two coupled single mode condensates provide one of the simplest examples to study this phenomenon in detail. 
In contrast to the case of small subsystems~ \cite{lightcone2,greiner1,bosonentanglement}, here the two coupled subsystems are equally large, and we find that equilibration can be  understood by approximating the state of the \emph{full} system by a \emph{microcanonical} ensemble rather than a thermal Gibbs ensemble.
%Since both subsystems are equally large, instead of forming a small subsystem with a large environment, 
%we expect that in this case the equilibration can be better studied in terms of a microcanonical ensemble instead of a thermal Gibbs ensemble. 
This system can be realized by loading a Bose-Einstein condensate into a double well potential (see Fig.~\ref{fig:intro0}). Assuming that the atoms in the left and right wells condense into a single wave function, the dynamics is governed by the Hamiltonian \cite{milburn}
\begin{equation}\label{eq:Hboson}
\hat{H}=-J\left(\hat{a}_L^\dagger\hat{a}_R+\hat{a}_R^\dagger\hat{a}_L\right)+\frac{U}{2}\left(\hat{N}_L^2-\hat{N}_L+\hat{N}_R^2-\hat{N}_R\right).
\end{equation}
Here the bosonic operators $\hat{a}_L^\dagger$ and $\hat{a}_R^\dagger$ create particles into the left and right potential wells respectively, and $\hat{N}_i=\hat{a}_i^\dagger\hat{a}_i$ for $i=L,R$. The first term in the Hamiltonian describes the tunneling of particles, while the second term takes into account the interaction between the bosons in the same potential well. For given total particle number $N$, the entanglement entropy between the left and right wells is simply given by \cite{entanglement}
\begin{equation}\label{eq:Sn}
\mathcal{S}(t)=-\sum_{n_L=0}^{N} P_t(n_L)\,\log P_t(n_L),
\end{equation}
where $P_t(n_L)$ denotes the probability of state $\hat{N}_L=n_L$ at time $t$ \cite{footnote0}. 
Here, concentrating on the effect of dephasing during the coherent, unitary time evolution of a closed quantum system,  
we investigate the time dependence of the entanglement entropy $\mathcal{S}(t)$ at $T=0$ temperature.
Importantly, while usually an entropy measurement would require detailed knowledge of a complicated quantum state, here the full time evolution of  $\mathcal{S}(t)$ can be investigated experimentally, since it only requires measuring the number of particles in the left and right wells. In contrast to earlier entropy measurements in optical lattices, involving a small sublattice with only a few atoms, coupled single-mode Bose-Einstein condensates would allow to study entanglement in large correlated many-body systems \cite{oberthalerentanglement,riedelentanglement,olsen}. 

Let us note that besides the double well experiment illustrated above, the Hamiltonian \eqref{eq:Hboson} can also be realized in a two component condensate trapped in a single well. E.g. two atomic hyperfine states forming the condensates may be coupled through microwaves \cite{hall}, while their interaction may be tuned using a Feshbach resonance \cite{feshbachrev}.

Hamiltonian \eqref{eq:Hboson} can also be rewritten in a more convenient form. Using the Schwinger boson representation~\cite{milburn},
 we introduce %the 
 spin operators 
\begin{equation*}
\hat{S}_z=\dfrac{1}{2}\left(\hat{N}_L-\hat{N}_R\right), \quad \hat{S}_x=\dfrac{1}{2}\left(\hat{a}_L^\dagger\hat{a}_R+\hat{a}_R^\dagger\hat{a}_L\right),
\end{equation*}
of length $N/2$, with $N$ denoting the total number of particles. Apart from a redundant constant term,  $\hat{H}$ can then be expressed as
\begin{equation}\label{eq:H}
\hat{H}= -2J\hat{S}_x+U\hat{S}_z^2.
\end{equation}
In this new representation, the entanglement entropy between the left and right wells corresponds to the entropy associated with  
$\hat{S}_z$ \cite{footnote},
\begin{equation}\label{eq:S}
\mathcal{S}(t)=-\sum_{m=-N/2}^{N/2} P_t(m)\,\log P_t(m)\;,
\end{equation}
with $P_t(m)$ denoting the probability of state $\hat{S}_z=m$ at time $t$. 

Let us note that the spin Hamiltonian Eq. \eqref{eq:H} is a special case of the Lipkin-Meshkov-Glick model, describing mutually interacting spin-1/2 particles, embedded in a magnetic field ~\cite{LMG}. In this context $\hat{S}_\alpha=\sum_i\hat{\sigma}_i^\alpha/2$ is the total spin operator, with  $\hat{\sigma}_i^\alpha$ denoting the Pauli matrices at site $i$ for $\alpha=x,y,z$. Depending on the the strength of the magnetic field, the Lipkin-Meshkov-Glick model shows a second order quantum phase transition. The entanglement properties of the ground state of this system have been analyzed by calculating the von Neumann entropy of a subsystem consisting of $L$ sites. In particular, it has been shown that the entanglement entropy shows a logarithmic divergence at the critical point of the quantum phase transition ~\cite{LMG1,QPT}. Similar divergence in the entanglement properties of the ground state at the critical point has also been observed in other systems, like the Dicke model or the transverse field Ising model ~\cite{dicke,fisherinfo}. Moreover, the dynamics of the von Neumann entropy of a single spin in the Lipkin-Meshkov-Glick model has also been investigated ~\cite{LMG2}. In this work we concentrate on a different type of entanglement entropy, associated with the spin operator $\hat{S}_z$.

\begin{figure}[b!]
\includegraphics[width=\columnwidth]{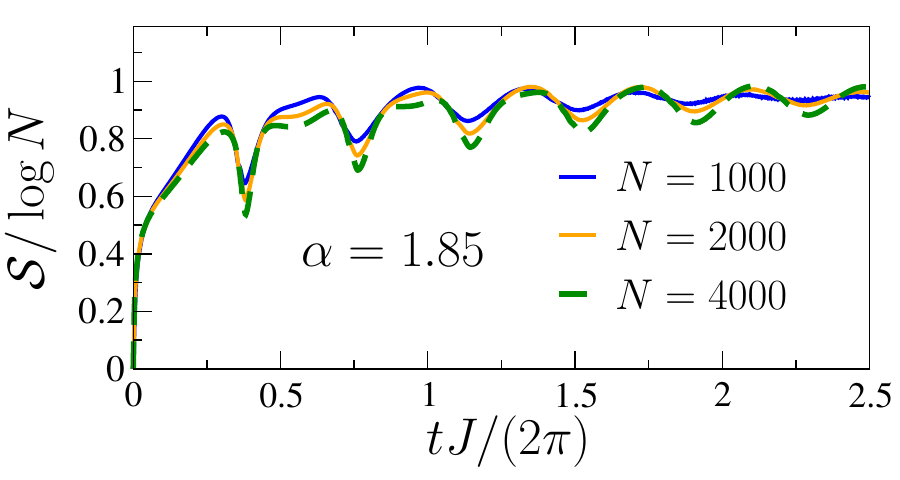}
\caption{Typical time evolution of entropy \eqref{eq:S}. Rescaled entropy $\mathcal{S}/\log N$ is plotted as a function of dimensionless time $t J/(2\pi)$ for different total particle numbers $N$, keeping $\alpha=1.85$ fixed. Initial state is chosen as $|\hat{S}_z=N/2\rangle$.  The entropy oscillates on the top of a steady increase, before saturating to a value proportional to $\log N$.}
\label{fig:intro2}
\end{figure}

Our main purpose here is to analyze the time evolution of entropy \eqref{eq:S} for different initial states and interaction strengths, by combining numerical results with analytical calculations. We demonstrate that $\mathcal{S}(t)$ exhibits coherent oscillations, reflecting the quantum mechanical dynamics of the coupled single-mode condensates. At the same time, $\mathcal{S}(t)$ shows a steady increase, and eventually reaches a stationary, "equilibrium" value, even though this closed system always remains in a pure state.

 The dynamics of the system depends crucially on the dimensionless parameter 
\begin{equation}\label{eq:alpha}
\alpha\equiv\dfrac{NU}{2J},
\end{equation}
characterizing the strength of interactions~\cite{dynamics}. For $\alpha<1$, the average population imbalance between the two potential wells, $N_L-N_R$, oscillates between positive and negative values. For $\alpha>1$, however, the system undergoes a self-trapping transition \cite{oberthaler,oberthalersemiclass}. Here, for large initial particle number imbalance $N_L-N_R$ and strong enough interactions $\alpha\gg 1$, the interaction energy of the initial state prevents levelling off the number of particles in the two wells, and the amplitude of population imbalance oscillations is suppressed (see Sec.~\ref{sec:semiclass} for more details).

We show a typical example of entropy production, e.g. steady increase of the entropy, in Fig~\ref{fig:intro2}. The initial state of the system corresponds to maximal population imbalance, $|\hat{S}_z=N/2\rangle$, and the time evolution of $\mathcal{S}$ is calculated numerically by exact diagonalization. The entropy shows oscillations on the top of a steady increase, before saturating to a constant value. Moreover, the curves corresponding to the same  $\alpha$, but different total particle numbers  can be scaled together. As we discussed earlier, these oscillations during entropy production should be experimentally accessible (for a discussion of experimental parameters see Sec. \ref{sec:microwave}). 

\begin{figure}[t!]
\includegraphics[width=0.75\columnwidth]{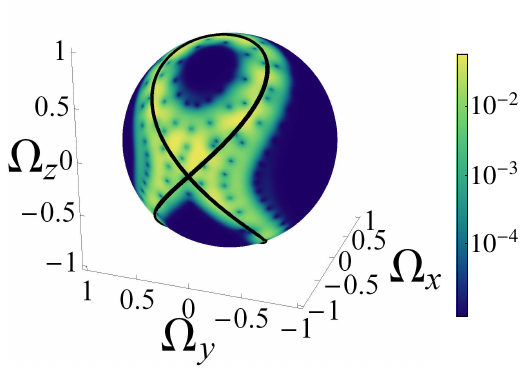}
\caption{Overlap between the wave function and different spin coherent states after dephasing. The overlap $|\langle\Omega_{\theta,\varphi}|\psi(t)\rangle|^2$ is plotted as a function of polar and azimuthal angles $\theta$ and $\varphi$ using logarithmic scale, ln, for interaction strength $\alpha=2$.  Here $|\Omega_{\theta,\varphi}\rangle$ is the spin coherent state of direction $(\theta,\varphi)$,  and $|\psi(t)\rangle$ denotes the wave function after time $tJ/(2\pi)=2.7$. We used the initial state $\hat{S}_z=N/2$ with $N=500$, lying on the boundary of self-trapping. Due to dephasing, the initial state quickly spreads over the vicinity of the classical trajectory (black line), allowing to apply a classical microcanonical description.}
\label{fig:introoverlap}
\end{figure}

The long time limit of the entropy also reflects the self-trapping transition by showing a sudden jump at the "phase boundary". Interestingly, 
the computed asymptotic entropy value agrees well with the predictions of a classical microcanonical ensemble, where the normalized spin vector $\vec{\Omega}\equiv 2\vec{S}/N$ is distributed uniformly along a classical trajectory. The remarkable success of classical description can be understood by investigating the overlap between the wave function $|\psi(t)\rangle$ and the spin coherent states $|\Omega_{\theta,\varphi}\rangle$, polarized into direction $(\sin\theta\cos\varphi,\sin\theta\sin\varphi,\cos\theta)$. We plotted this overlap on the unit sphere in Fig.~\ref{fig:introoverlap}, for a maximally polarized initial state $\hat{S}_z=N/2$, with the interaction strength tuned to the boundary of self-trapping transition. For sufficiently large $t$, the dephasing between different energy eigenstates leads to the broadening of the wave function, and the overlap $|\langle\Omega_{\theta,\varphi}|\psi(t)\rangle|^2$ traces out precisely the semiclassical trajectory (black line in Fig.~\ref{fig:introoverlap}).

The paper is organized as follows. In Sec. \ref{sec:semiclass}, we outline the semi-classical dynamics of Hamiltonian \eqref{eq:H} \cite{phasespace,semiclass,semiclass2}. We analyze the entropy oscillations and the entropy production for short times, and compare the exact dynamics to a semi-classical approximation in Sec. \ref{sec:shortt}. In Sec. \ref{sec:longt} we concentrate on the stationary long time limit of the entropy, and show that it is well approximated by the classical entropy of the microcanonical ensemble. We outline the experimental realization of Hamiltonian \eqref{eq:H} in microwave measurements with $^{87}$Rb atoms in Sec. \ref{sec:microwave}. Our conclusions are summarized in Sec. \ref{sec:discuss}.

\section{Semiclassical dynamics}\label{sec:semiclass}

\begin{figure}[b!]
\includegraphics[width=\columnwidth]{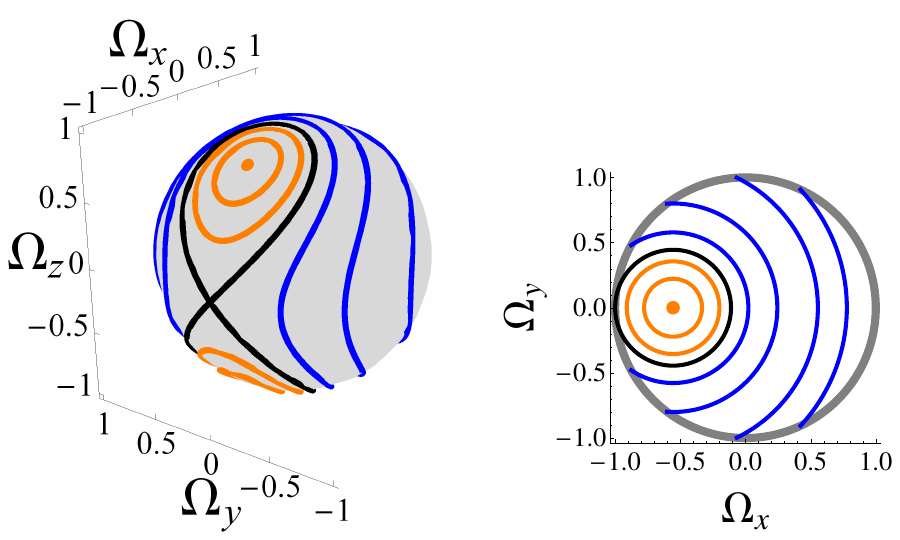}
\caption{Left: Unit sphere, and classical spin trajectories above the self-trapping transition $\alpha>\alpha_c$. Trapped trajectories (orange) never cross the equator, but remain confined to the upper or lower hemisphere. Non-trapped trajectories (blue) intersect the equator, visiting both hemispheres. The boundary of these regimes is the separatrix (black). Right: Projection of trajectories to the $x-y$ plane. Trapped trajectories (orange) form full circles, while non-trapped curves (blue) are arc segments inside the unit circle (grey). The separatrix (black line) touches the unit circle at $\Omega_x=-1$.}
\label{fig:semiclass}
\end{figure}

In the following sections we will investigate the time evolution of the entropy \eqref{eq:S}, with different spin coherent states taken as initial conditions. More precisely, we consider spin coherent states lying in the $x-z$ plane, polarized in the direction $\vec{\Omega}_\theta=(\sin\theta,\,0,\,\cos\theta)$, for different angles $-\pi/2<\theta<\pi/2$. These states are eigenstates of the spin operator $\hat{S}_\theta=\hat{S}_x\sin\theta+\hat{S}_z\cos\theta$, with eigenvalue $\hat{S}_\theta=N/2$.  For large total particle number $N$, the semiclassical approximation yields a good description \cite{semiclass,oberthalersemiclass,semiclass2}, and the spin operators in Eq. \eqref{eq:H} can be replaced by the components of a classical vector
\begin{equation*}
\vec{S}=\dfrac{N}{2}\vec{\Omega}.
\end{equation*}
The time evolution of the unit vector $\vec{\Omega}$ is governed by the differential equations \cite{semiclass}
\begin{align}\label{eq:classdyn}
&\partial_t\Omega_x=-UN\Omega_y\Omega_z,\nonumber\\
&\partial_t\Omega_y=2J\Omega_z+UN\Omega_x\Omega_z,\nonumber\\
&\partial_t\Omega_z=-2J\Omega_y,
\end{align}
with initial condition $\vec{\Omega}(t=0)=\vec{\Omega}_\theta$. These classical trajectories lie on the unit sphere, and their shape is determined by the parameter $\alpha$ in Eq. \eqref{eq:alpha} \cite{semiclass}. 

Typical trajectories are depicted in Fig.~\ref{fig:semiclass}. Below the critical value $\alpha_c=1$, all trajectories intersect the equator of the sphere, and no self-trapping occurs. Here the equations of motion \eqref{eq:classdyn} have two stable fixed points at $\Omega_x=\pm 1$, $\Omega_y=\Omega_z=0$. For $\alpha>\alpha_c$, however, a self-trapped regime appears on the unit sphere (see Fig.~\ref{fig:semiclass}). Here the fixed point at $\Omega_x=-1$ becomes unstable, and bifurcates into two new, stable fixed points at \cite{semiclass,oberthalersemiclass}
\begin{equation*}
\Omega_x=-\dfrac{1}{\alpha},\quad\Omega_y=0,\quad\Omega_z=\pm\sqrt{1-\dfrac{1}{\alpha^2}}.
\end{equation*}
Trapped trajectories around these fixed points can not cross the equator of the sphere, but are constrained to the $\Omega_z>0$ or $\Omega_z<0$ hemisphere. Non-trapped trajectories, however, reach both positive and negative $\Omega_z$ values. The separatrix, forming the boundary of self-trapping, touches the equator at the unstable fixed point, $\Omega_x=-1$ (see Fig.~\ref{fig:semiclass}).

The semiclassical trajectories can be visualized more easily by noting that their projections on the $x-y$ plane form circles centered at $(-1/\alpha,0)$ (see Fig.~\ref{fig:semiclass}),
\begin{equation}\label{eq:classtrajectory}
\left(\Omega_x+\dfrac{1}{\alpha}\right)^2+\Omega_y^2={\rm const.}
\end{equation}
 The trajectory determined by the initial condition $\vec{\Omega}(t=0)=(\sin\theta,0,\cos\theta)$ will coincide with the separatrix at interaction strength
\begin{equation}\label{eq:separatrix}
\alpha_{\theta}=\dfrac{2}{1-\sin\theta}.
\end{equation}
Another special case occurs, when the initial condition satisfies
\begin{equation}\label{eq:fixpoint}
\alpha_{\theta}^{\rm fix}=-\dfrac{1}{\sin\theta},
\end{equation}
and $\vec{\Omega}(t=0)$ is a stable fixed point of the classical equations of motion \eqref{eq:classdyn}. 

As we will show later, the long time limit of the entropy \eqref{eq:S} reflects this semiclassical dynamics (see Sec. \ref{sec:longt}). The trapping transition at $\alpha=\alpha_\theta$, Eq. \eqref{eq:separatrix}, is revealed by a sudden jump of size $\log 2$ in the entropy, related to the rapid change by a factor of $2$ in the length of the classical trajectory. The classical fixed point \eqref{eq:fixpoint} corresponds to a local minimum in $\mathcal{S}$ due to the strong confinement of trajectories around this point.

\section{Entropy production}\label{sec:shortt}

Now we concentrate on the entropy production at short times, during the first few oscillations of entropy \eqref{eq:S}. We consider the spin coherent initial state $|\Omega_\theta\rangle\equiv|\hat{S}_\theta=N/2\rangle$, with $\hat{S}_\theta=\hat{S}_x\sin\theta+\hat{S}_z\cos\theta$. To gain more insight into the structure of the wave function, let us expand this state in the eigenbasis of $\hat{S}_z$ \cite{Auerbach},
\begin{align*}
&|\Omega_\theta\rangle=\\
&\sum_{m=-N/2}^{N/2}\sqrt{\binom{N}{m+\frac{N}{2}}}\left(\cos\frac{\theta}{2}\right)^{m+\frac{N}{2}}\left(\sin\frac{\theta}{2}\right)^{\frac{N}{2}-m}|m\rangle,
\end{align*}
with $|m\rangle$ denoting the eigenstate $\hat{S}_z=m$. This expression shows that the shifted spin operator $\hat{S}_z+N/2$ follows a binomial distribution $B(n,p)$, with number of trials $n=N$ and probability $p=\cos^2(\theta/2)$. This binomial distribution yields an expectation value $\langle\hat{S}_z\rangle=np-N/2=N/2\cos\theta$ and a variance ${\rm Var}(\hat{S}_z)=np(1-p)=N\sin^2\theta/4$. In the semiclassical limit of large total particle number $N$, this initial state can be approximated by a Gaussian wave function
\begin{equation}\label{eq:gaussinitial}
|\Omega_\theta\rangle\approx\sqrt{\frac{2}{N\pi\sin^2\theta}}\sum_m\exp\left(-\dfrac{\left(m-\frac{N}{2}\cos\theta\right)^2}{N\sin^2\theta}\right)|m\rangle,
\end{equation}
excepting the vicinity of $\theta=0$. 

\begin{figure}[t!]
\includegraphics[width=\columnwidth]{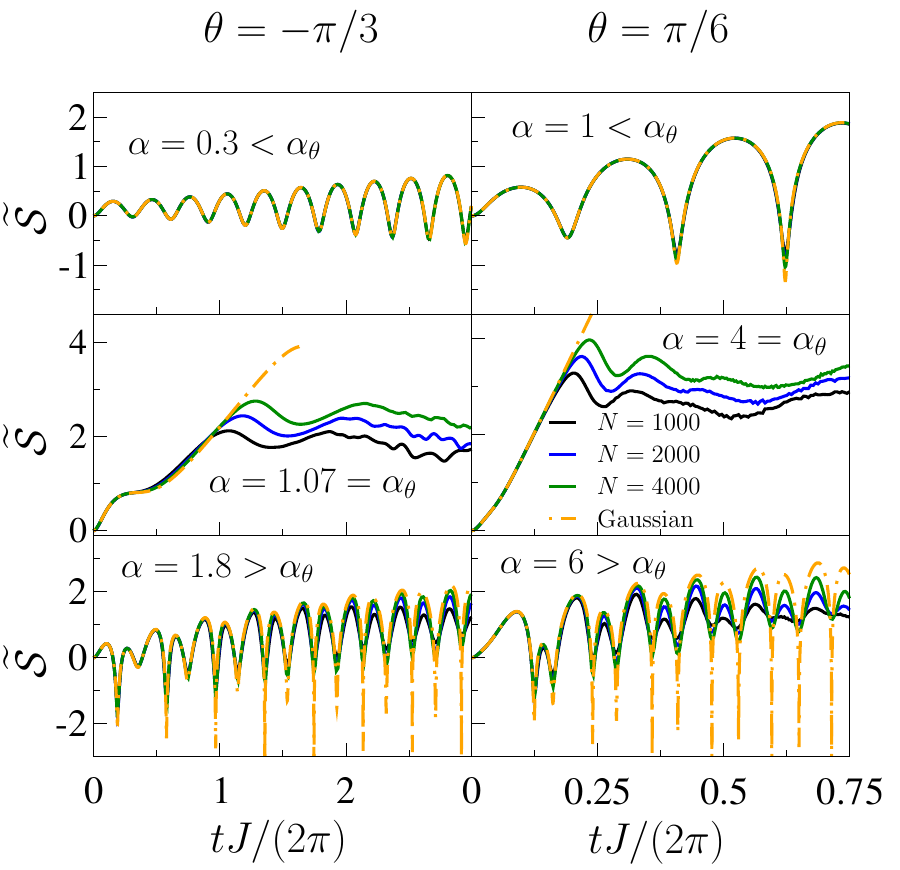}
\caption{Entropy production for short times. Time evolution of rescaled entropy $\widetilde{\mathcal{S}}$ is plotted as a function of dimensionless time $t\,J/(2\pi)$, for different interaction strengths $\alpha$ and initial conditions $\theta$. Different rows correspond to non-trapped regime (top), phase boundary $\alpha=\alpha_\theta$ (center) and self-trapping (bottom). Curves with different particle numbers $N$, shown in different colors, scale together for short times, before they reach a stationary value differing by $\log\sqrt{N}$. On the top of the steady increase of entropy, $\widetilde{\mathcal{S}}$ oscillates both in the non-trapped and self-trapped regimes. These oscillations vanish at the phase boundary, leaving an approximately linear increase of the entropy. The analytical results of a Gaussian, semiclassical approximation (dashed orange line), Eq. \eqref{eq:SG}, are also shown.}
\label{fig:shorttime}
\end{figure}

The entropy of a Gaussian distribution is known exactly \cite{Renyi}, yielding the approximation
\begin{equation*}
\mathcal{S}(t=0)\approx\dfrac{1}{2}\log\dfrac{\pi\,eN\sin^2\theta}{2}.
\end{equation*}
Based on this expression, we scale together the curves corresponding to different total particle numbers $N$ by introducing the rescaled entropy
\begin{equation}\label{eq:rescale}
\widetilde{\mathcal{S}}(t)=\mathcal{S}(t)-\dfrac{1}{2}\log\dfrac{\pi\,eN\sin^2\theta}{2}.
\end{equation}

The time evolution of the rescaled entropy \eqref{eq:rescale}, obtained by exact diagonalization, is shown for different interactions strengths $\alpha$ and particle numbers $N$ in Fig.~\ref{fig:shorttime}, for two different initial angles $\theta$. Different rows correspond to the non-trapped regime $\alpha<\alpha_\theta$ (top), lie on the phase boundary $\alpha=\alpha_\theta$ (center), and fall in the self-trapped regime $\alpha>\alpha_\theta$ (bottom), respectively, with $\alpha_\theta$ given by Eq. \eqref{eq:separatrix}. As expected, for fixed $\alpha$ but different total particle number $N$, the rescaled entropies $\widetilde{\mathcal{S}}$ follow the same curve for short times. The long time limit of $\widetilde{\mathcal{S}}$, however, is proportional to $\log N$ instead of $\log\sqrt{N}$, and is different for each $N$ (see also Fig.~\ref{fig:intro2}). The entropy oscillates both in the non-trapped and self-trapped regimes, while increasing steadily towards a stationary value. These entropy oscillations reveal the coherent oscillations of the single mode condensates, while the dephasing between different energy eigenstates is responsible for the steady increase of the entropy. At the phase boundary, the oscillations become washed out, and the entropy increases approximately linearly, until saturating to the long time limit. 

The main features of the time evolution of $\mathcal{S}$ can be understood in terms of the classical trajectories in the semiclassicl limit of large $N$. As supported by the detailed analysis below, the state of the system can be visualized as an extended packet on the unit sphere around the classical unit vector $\vec{\Omega}(t)$. For an initial state $\hat{S}_\theta=N/2$, this packet has a Gaussian shape around $\vec{\Omega}_\theta$, with variance $\sim 1/N$. The center of the packet, $\vec{\Omega}(t)$, follows the classical equations of motion \eqref{eq:classdyn}, while the typical width of the packet increases. This broadening occurs because the packet gets more elongated along the classical trajectory due to the dephasing between different energy eigenstates. At the same time, the width perpendicular to the trajectory decreases to keep the volume of the packet constant.  

The distribution of $\Omega_z$ and the corresponding entropy $\mathcal{S}$ can be determined by projecting this packet to the $z$ axis. The oscillations of the center of the packet, $\vec{\Omega}(t)$, result in entropy oscillations, and their period is given by the period of the classical trajectory. For the separatrix this period is infinity, explaining the vanishing entropy oscillations at the boundary of self-trapping. For a more detailed analysis, notice that the entropy is proportional to $\log \sigma$, with $\sigma$ denoting the typical width of the distribution of $\Omega_z$. In spite of the broadening of the packet, $\sigma$ can display a very different behavior depending on the position along the trajectory. At the upper and lower turning points, where the tangent vector of the trajectory is
perpendicular to the axis $\hat z$, the projection yields a sharp distribution for $\Omega_z$, resulting in local minima for the entropy. Since the width of the packet perpendicular to the trajectory decreases, $\sigma$ can even decrease compared to the width of the initial state, resulting in decreasing local minima (see the first row in Fig.\ref{fig:shorttime}). On the other hand, at the horizontal turning points, where the tangent vector is parallel to $\hat z$, 
 $\sigma$ is maximal. This maximal value  increases with time as the wave  packet gets more elongated along the trajectory, yielding increasing entropy
  maxima after each oscillation. 

To substantiate these arguments and to get a quantitative description for the time evolution of entropy $\widetilde{\mathcal{S}}$, we applied a Gaussian Ansatz for the wave function. This approximation relies on the observation that the initial state is well described by the Gaussian expression \eqref{eq:gaussinitial}. We assume that the wave function keeps this Gaussian form during the time evolution. As a first step, we expand the wave function according to the eigenstates of $\hat{S}_z$, 
$$|\psi(t)\rangle=\sum_m e^{-i\varphi(t)m}\, c_m(t)\,|m\rangle .$$
Having separated a rapidly oscillating phase factor $e^{-i\varphi(t)m}$ - corresponding to the rotation of the state around the $z$ axis - we can assume that the coefficients $c_m(t)$ are slowly varying functions of $m$.

Let us introduce a new variable $x=2m/N$. In the limit of large $N$, $x$ can be treated as a continuous variable \cite{WKB}. We can replace the discrete, slowly varying coefficients $c_m(t)$ by a continuous function $\psi(x,t)$, and assume a Gaussian form,
\begin{align}\label{eq:gaussian}
&c_m(t)\rightarrow \psi(x,t)\equiv  \nonumber\\ 
& \left(\frac{2N{\rm Re }\,c(t)}{\pi}\right)^{1/4}\exp(-c(t)N\, (x-x_0(t))^2).
\end{align}
This Ansatz yields a Gaussian distribution for the normalized spin operator $\hat{\Omega}_z\equiv 2\hat{S}_z/N$, with expectation value $\langle\hat{\Omega}_z\rangle(t)=x_0(t)$ and variance $1/(N{\rm Re}\,c(t))$. Moreover, for the Gaussian wave function given by Eq. \eqref{eq:gaussian}, the rescaled entropy \eqref{eq:rescale} can be expressed as 
\begin{equation}\label{eq:SG}
\mathcal{S}_G(t)=-\dfrac{1}{2}\log\left(4\sin^2\theta\;{\rm Re}\,c(t)\right).
\end{equation} 

The optimal parameters of the Gaussian wave function, $|\psi_G\rangle$, are determined from the variational condition
\begin{equation}\label{eq:variational}
\delta\langle\psi_G|\,i\partial_t-\hat{H}\,|\psi_G\rangle=0,
\end{equation}
where
\begin{align*}
&\langle\psi_G|\,i\partial_t-\hat{H}\,|\psi_G\rangle=\\
&\partial_t\varphi(t)\,\frac{N}{2}\int{\rm d}x\,x\,|\psi(x,t)|^2+i\int{\rm d}x\,\psi^*(x,t)\,\partial_t\psi(x,t)\\
&-U\frac{N^2}{4}\int{\rm d}x\,x^2\,|\psi(x,t)|^2 +\dfrac{JN}{2}\left(e^{i\varphi(t)}\right.\times\\
&\left.\int {\rm d}x\,\psi^*(x,t)\psi(x-\frac{2}{N},t)\sqrt{1-x^2+\frac{2}{N}(1+x)}+c.c.\right).\end{align*}
For large total particle number $N$, Eq. \eqref{eq:variational} can be expanded systematically according to the powers of $N$. The leading order contributions result in the semiclassical equations of motion
\begin{align}\label{eq:xphi}
&\partial_t x_0=-2J\sqrt{1-x_0^2}\,\sin\varphi,\nonumber\\
&\partial_t\varphi=UNx_0+2J\dfrac{x_0}{\sqrt{1-x_0^2}}\cos\varphi.
\end{align}
These equations determine the same trajectories as Eqs. \eqref{eq:classdyn}, with the unit vector $\vec{\Omega}$ given by
\begin{equation*}
\vec{\Omega}=(\sqrt{1-x_0^2}\,\cos\varphi,\,\sqrt{1-x_0^2}\,\sin\varphi,\,x_0).
\end{equation*}
The next order of the expansion yields the time evolution of $c(t)$,
\begin{align*}
& i\partial_t c=-\dfrac{\alpha J}{2}-\dfrac{J\cos\varphi}{2\,(1-x_0^2)^{3/2}}\\
&\quad\quad-4J\dfrac{x_0}{\sqrt{1-x_0^2}}\sin\varphi\;c+8J\cos\varphi\sqrt{1-x_0^2}\;c^2.
\end{align*}
Notice that $c(t)$ only depends on the dimensionless time $tJ$, the parameter $\alpha$ and the initial condition $\theta$, but not on the particle number $N$. Concentrating on the semiclassical limit of large $N$, we neglect the remaining $O(1/N)$ corrections.

The Gaussian entropy $\eqref{eq:SG}$ is plotted together with exact numerical results in Fig.~\ref{fig:shorttime}. As noted above, $\mathcal{S}_G$ is independent of $N$ up to corrections of the order $1/N$, neglected in our semiclassical approximation. We find that Eq. \eqref{eq:SG} yields a surprisingly good approximation for the dynamics at short times.

Let us emphasize that the Gaussian ansatz relied on the observation that the spin coherent initial state results in a Gaussian distribution for $\hat{S}_z$, thus this description remains valid only on time scales shorter than the time scale of the entropy saturation. Indeed, in Fig.~\ref{fig:shorttime} we have found that a Gaussian wave function remains a reasonable approximation on such short time scales. However, as we show in Sec. \ref{sec:longt} below, this ansatz breaks down as the entropy saturates to the stationary long time limit (see also the middle row of Fig.~\ref{fig:shorttime}). On such long time scales it has to be replaced by a non-Gaussian semiclassical approximation, derived in Sec. \ref{sec:longt} and Appendix \ref{app:PSz}.

\section{Long time limit of entropy and equilibration}\label{sec:longt}

We now turn to the long time behavior of the entropy $\mathcal{S}$, and show how it reflects the semiclassical dynamics discussed in Sec. \ref{sec:semiclass}. Due to the discrete spectrum of Hamiltonian \eqref{eq:H}, the entropy shows several revivals and, strictly speaking, it never reaches a stationary value. However, the period of these revivals is typically very long compared to experimentally relevant time scales, and it is still meaningful to consider the steady state at intermediate times \cite{revival}.

\begin{figure}[t!]
\includegraphics[width=\columnwidth]{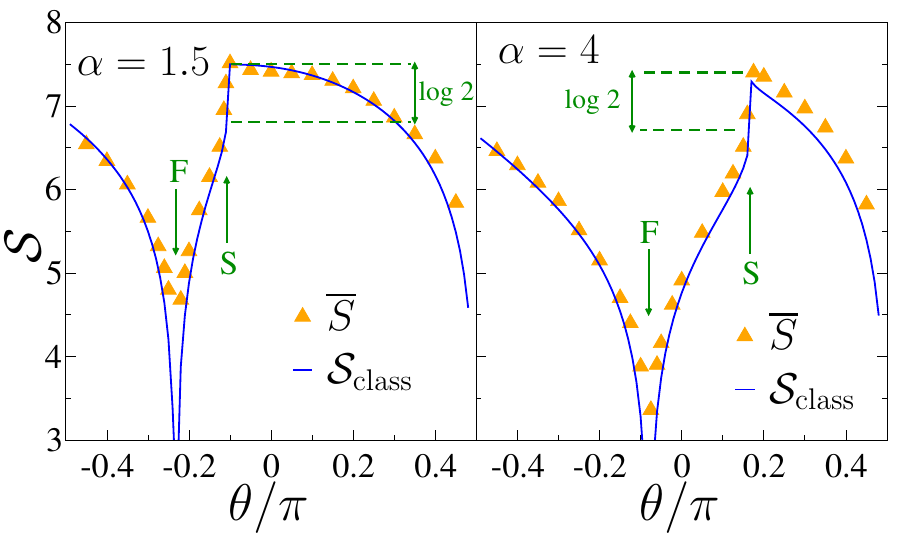}
\caption{Long time limit of entropy. Time averaged entropy $\overline{S}$ (symbols) plotted as a function of initial angle $\theta$, for two different parameters $\alpha$. Semiclassical fixed point (F), Eq. \eqref{eq:fixpoint}, appears as a sharp minimum in the entropy. The separatrix of self-trapping (S) is accompanied by a sudden entropy gain of size $\log 2$, due to the doubling of available phase space at the phase transition. The prediction of a semiclassical microcanonical ensemble (solid line) is also shown. For the numerics we used total particle number $N=3000$.}
\label{fig:longtime}
\end{figure}

We defined the long time limit of the entropy \eqref{eq:S} as the time average
\begin{equation}\label{eq:Stav}
\overline {\mathcal{S}}=\dfrac{1}{T}\int_0^T {\rm d}t\,\mathcal{S}(t),
\end{equation}
with $T$ chosen large enough to reach a stationary value. As before, we used spin coherent initial states $|\hat{S}_\theta=N/2\rangle$. The numerical results from exact diagonalization are shown in Fiq.~\ref{fig:longtime} as a function of the angle $\theta$, for two different parameters $\alpha$. 

The semiclassical fixed point, Eq. \eqref{eq:fixpoint}, appears as a sharp local minimum in the time averaged entropy. Since the entropy is related to the width of the distribution of $\hat{S}_z$, this entropy minimum follows from the strong confinement of classical trajectories around the stable fixed point, leading to sharp distributions for $\Omega_z$. The separatrix of the self-trapped phase, Eq. \eqref{eq:separatrix}, is revealed by a sudden jump of size $\log 2$ in entropy \eqref{eq:Stav}. This sudden entropy gain is related to the doubling of the length of classical trajectories at the self-trapping transition, doubling the phase space available for $\vec{\Omega}$ (see Fig.~\ref{fig:semiclass}).

The long time limit of the entanglement entropy, Eq. \eqref{eq:Stav}, can be understood in terms of a semiclassical microcanonical ensemble. In the semiclassical approximation, the trajectory is determined by Eqs. \eqref{eq:classdyn}, which conserves the energy of the classical Hamiltonian. In a microcanonical description, the spin vector $\vec{\Omega}$ is randomly distributed along this trajectory (the surface of constant energy in general), resulting in a uniform distribution on the classical trajectory. This classical trajectory amounts in a continuous distribution for the $z$-component of the spin, $\Omega_z$ (see Appendix \ref{app:PSz}). Denoting the corresponding probability density by $P(\Omega_z)$, the classical entropy is given by
\begin{equation}\label{eq:Sclass}
\mathcal{S}_{\rm class}=-\int\,{\rm d}\Omega_z P(\Omega_z)\log P(\Omega_z)+\mathcal{S}_0.
\end{equation}
Here $\mathcal{S}_0$ denotes an arbitrary constant entropy shift, accounting for some unknown box size $\Delta\Omega_z$. 
%\li{Let us emphasize again that the classical distribution $P(\Omega_z)$ (derived in Appendix A) is crucially different from the Gaussian distribution considered in Sec. \ref{sec:shortt}. The Gaussian ansatz breaks down as the entropy reaches the stationary long time limit and the wave function extends over the full vicinity of the classical trajectory (see the middle row of Fig.~\ref{fig:shorttime}). Instead, the non-Gaussian semiclassical approximation ~\eqref{eq:Sclass} is valid here.}
Eq.~\eqref{eq:Sclass} yields good agreement with the numerical results by using a single fitting parameter $\mathcal{S}_0=7.0$ (see Fig.~\ref{fig:longtime}). The deviation between the semiclassical approximation and $\overline{\mathcal{S}}$ gets larger only in the immediate vicinity of the semiclassical fixed point, Eq. \eqref{eq:fixpoint}. Here $\mathcal{S}_{\rm class}$ diverges, because the variance of the continuous classical distribution $P(\Omega_z)$ approaches zero. However, the time averaged entropy $\overline{\mathcal{S}}$ remains non-negative even at the fixed point, and its minimal value is determined by the width of the spin coherent initial state in the eigenbasis of $\hat{S}_z$. For the semiclassical case of large $N$ this yields a minimal variance ${\rm Var}(\hat{S}_z)\approx N/(4\,\alpha^2)$ and minimal entropy
\begin{equation*}
\overline{\mathcal{S}}_{\rm min}\approx\dfrac{1}{2}\log\dfrac{\pi\,e\,N}{2\, \alpha^2}.
\end{equation*}

\begin{figure}[t!]
\includegraphics[width=\columnwidth]{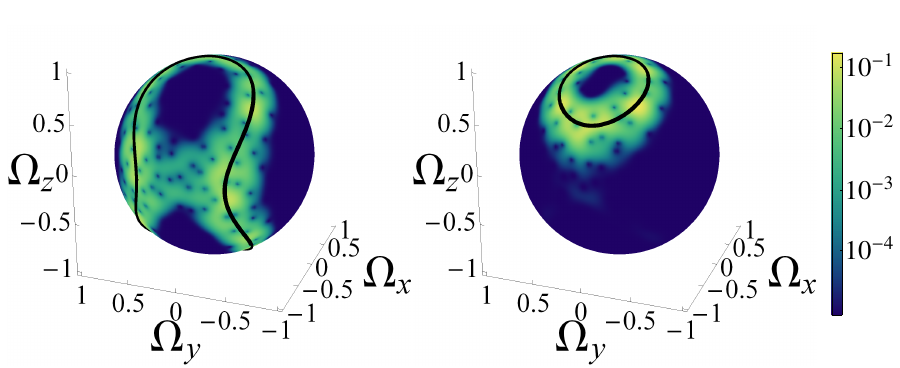}
\caption{Overlap between the wave function and different spin coherent states for long times. The overlap $|\langle\Omega_{\theta,\varphi}|\psi(t)\rangle|^2$ is plotted as a function of polar and azimuthal angles $\theta$ and $\varphi$ using logarithmic scale, ln, for two different interaction strengths corresponding to non-trapped ($\alpha=1.8$, left) and self-trapped ($\alpha=2.4$, right) regimes, respectively. Here $|\Omega_{\theta,\varphi}\rangle$ denotes the spin coherent state of direction $(\theta,\varphi)$. We used the maximally polarized initial state $\hat{S}_z=N/2$ with $N=500$, and $|\psi(t)\rangle$ is the wave function after time $tJ/(2\pi)=19.3$. Classical trajectories (black lines) are also shown for comparison. During the time evolution the spin coherent initial state broadens and becomes elongated along the classical trajectory.}
\label{fig:overlap}
\end{figure}

\begin{figure}[t!]
\includegraphics[width=\columnwidth]{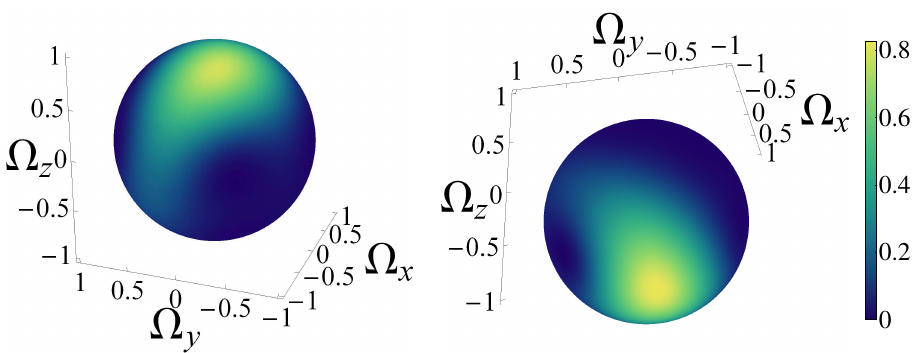}
\caption{Tunneling process on long time scales. The overlap $|\langle\Omega_{\theta,\varphi}|\psi(t)\rangle|^2$ is plotted as a function of polar and azimuthal angles $\theta$ and $\varphi$ at two different times $t$, for a small system with total particle number $N=10$. Here $|\Omega_{\theta,\varphi}\rangle$ denotes the spin coherent state of direction $(\theta,\varphi)$, and $|\psi(t)\rangle$ is the wave function at time $t$. We used the maximally polarized initial state $\hat{S}_z=N/2$ in the self-trapped regime with $\alpha=3$. At intermediate times ($tJ=110$, left) the particles remain trapped due to the large interaction energy of the initial state, whereas on longer time scales ($tJ=900$, right) the system can tunnel to reversed population imbalances.}
\label{fig:tunnel}
\end{figure}

A classical microcanonical equilibrium ensemble thus yields  a good approximation for the time averaged entropy, in spite of the unitary time evolution of the system. To see how dephasing alone can lead to equilibration, we plotted the overlap between the wave function $|\psi(t)\rangle$ and the spin coherent states polarized into direction $(\sin\theta\cos\varphi,\sin\theta\sin\varphi,\cos\theta)$, $|\Omega_{\theta,\varphi}\rangle$,  for long times in Fig.~\ref{fig:overlap}. For the  initial state we chose the maximally polarized state $\hat{S}_z=N/2$, and selected two different interaction strengths, corresponding the non-trapped and self-trapped regimes, respectively. In accordance with the classical picture presented above, the wave function broadens, and extends to the vicinity of the classical trajectory on both sides of the self-trapping transition. Note that in spite of the success of this semiclassical description, Fig.~\ref{fig:overlap} still reflects the non-classical nature of the exact wave function, by displaying several very sharp minima on the unit sphere, originating from quantum interference. Notice also that the state of the system  never becomes stationary, and at any time the density displays  several maxima along the trajectory. The position of these maxima depends on time, leading to revival effects. However, the time averaged entropy is always well approximated by a state spreaded uniformly along the classical trajectory.

As a final remark, let us note that in order to observe self-trapping, the upper bound of the time averaging Eq.~\eqref{eq:Stav}, $T$, should not be too large. Strictly speaking, a real self-trapping transition only occurs in the classical system. The quantum system would eventually tunnel to reversed population imbalances \cite{tunneling}, and the entropy would increase by $\log 2$, compensating the entropy loss of self-trapped regime. We have chosen $T$ much smaller than the time scale of this tunneling process, $T_{\rm tunnel}$. The period $T_{\rm tunnel}$ depends on the small energy difference $\Delta E_{\rm doublet}$ between quasi-degenerate doublets $(|u\rangle\pm|d\rangle)/\sqrt{2}$, with the states $|u\rangle$ and $|d\rangle$ confined to the $\hat{S}_z>0$  and $\hat{S}_z<0$ hemispheres, respectively. The splitting $\Delta E_{\rm doublet}$ is tiny even for moderate particle numbers ~\cite{footnote2}, resulting in an exponentially  long tunneling time  $T_{\rm tunnel}$, far beyond  the time scales available for experiments or numerical simulations. However, for
small  systems consisting only of a few particles, this tunneling process can be observed in simulations. An example of this population inversion is shown in Fig.~\ref{fig:tunnel}, for a small particle number $N=10$\li{ ~\cite{footnote4}}.

\section{Microwave experiments}\label{sec:microwave}

As mentioned in the introduction, Hamiltonian \eqref{eq:H} can also be realized in microwave experiments with ultracold atoms, making use of different hyperfine states of the atoms. Below we outline how Hamiltonian \eqref{eq:H} arises in this setup, and we briefly discuss the optimal experimental parameters for the observation of entropy oscillations for the specific case of $^{87}$Rb atoms.

In $^{87}$Rb experiments, one can tune interactions by utilizing the two hyperfine states $|0\rangle\equiv |F=1,\,m_F=1\rangle$ and $|1\rangle\equiv |F=2,\,m_F=-1\rangle$, with $F$ and $m_F$ denoting the total spin of the atom and its projection to the quantization axis, respectively \cite{feshbachrev}. These states can be trapped in optical dipolar traps, while they can be coupled by microwave pulses. Below we concentrate on an experimentally relevant setting, where these atoms are trapped in a spherically symmetric harmonic trap with trapping frequency $\nu_0=50$Hz. For weak enough interactions, all atoms occupy the ground state of this harmonic potential,
\begin{equation}\label{eq:harmonic}
\varphi_0(\mathbf{r})=\dfrac{1}{\pi^{3/4}l_0^{3/2}}\exp\left(-\dfrac{r^2}{2l_0^2}\right),
\end{equation}
with $l_0=\sqrt{\hbar/(2\pi m\nu_0)}$ denoting the oscillator length. Let us note that for stronger interactions the condensate wave function is better described by a Thomas-Fermi profile instead of $\varphi_0$, Eq. \eqref{eq:harmonic}. However, for not too large interaction strengths, a two mode approximation is still applicable, and the derivation presented below is valid with minor modifications (see also the discussion at the end of the section).

The short range interaction between the Rb atoms is well described by a Dirac-delta potential. Denoting the bosonic creation operators of the hyperfine states $|0\rangle$ and $|1\rangle$ by $\hat{a}_0^\dagger$ and $\hat{a}_1^\dagger$, the interaction energy is given by
\begin{align}\label{eq:Hint}
\hat{H}_{\rm int}&=\sum_{\sigma,\sigma^\prime=0,1}\dfrac{g_{\sigma\sigma^\prime}}{2}\int{\rm d}^3\mathbf{r}\,|\varphi_0(\mathbf{r})|^4\,\hat{a}_\sigma^\dagger\hat{a}_{\sigma^\prime}^\dagger\hat{a}_{\sigma^\prime}\hat{a}_\sigma\nonumber\\
&=\sum_{\sigma,\sigma^\prime=0,1}\dfrac{U_{\sigma\sigma^\prime}}{2}\hat{a}_\sigma^\dagger\hat{a}_{\sigma^\prime}^\dagger\hat{a}_{\sigma^\prime}\hat{a}_\sigma.
\end{align}
Here $U_{\sigma\sigma^\prime}=g_{\sigma\sigma^\prime}/(2\pi l_0^2)^{3/2}$, and the interaction strength $g_{\sigma\sigma^\prime}$ can be expressed with the scattering length of the Rb atoms, $a_{\sigma\sigma^\prime}$, as \cite{ZwergerReview}
\begin{equation*}
g_{\sigma\sigma^\prime}=\dfrac{4\pi\hbar^2}{m}a_{\sigma\sigma^\prime},
\end{equation*}
with $m$ denoting the mass of $^{87}$Rb.

The bare scattering lengths of $^{87}$Rb depend very weakly on the hyperfine states of the atoms, and all interactions are determined by the single length scale $a_{\sigma\sigma^\prime}=5.3$nm. The scattering length between hyperfine states $|0\rangle$ and $|1\rangle$, however, can be tuned by a Feshbach resonance, changing $a_{01}$ by as much as $\Delta a_{01}=0.1\, a_{00}=0.53$nm \cite{feshbach}. Introducing the average interaction strength $\overline{U}=(U_{00}+U_{01})/2$, and the difference $\Delta U=U_{00}-U_{01}$, the interaction energy \eqref{eq:Hint} can be rewritten as
\begin{align*}
\hat{H}_{\rm int}&=\dfrac{U_{00}}{2}(\hat{N}_0^2-\hat{N}_0+\hat{N}_1^2-\hat{N}_1)+U_{01}\hat{N}_0\hat{N}_1\\
&=\dfrac{\overline{U}}{2}N^2+\Delta U\hat{S}_z^2.
\end{align*} 
Here $\hat{N}_i=\hat{a}_i^\dagger\hat{a}_i$ for $i=0,1$, $N=\hat{N}_0+\hat{N}_1$ is the total particle number, and the spin operator is defined as $\hat{S}_z=(\hat{N}_0-\hat{N}_1)/2$. For a closed system $\overline{U}N^2/2$ is just an irrelevant constant energy shift. Thus the interaction between Rb atoms takes the same form as the interaction term in Hamiltonian \eqref{eq:H}, with interaction strength $\Delta U$ determined by the difference of scattering lengths $a_{00}-a_{01}$.

Let us mention that instead of controlling the scattering lengths $a_{\sigma\sigma^\prime}$ by a Feshbach resonance, the interaction strength $\Delta U$ can also be tuned by applying a microwave trapping potential which depends on the hyperfine state of the atoms \cite{riedelentanglement}. In this case the atoms occupy state-dependent condensate wave functions, $\varphi_\sigma$, and the interaction strength is given by
\begin{equation*}
U_{\sigma\sigma^\prime}=g_{\sigma\sigma^\prime}\int{\rm d}^3\mathbf{r}\,|\varphi_\sigma(\mathbf{r})|^2|\varphi_{\sigma^\prime}(\mathbf{r})|^2.
\end{equation*}
Thus $\Delta U=U_{00}-U_{01}$ can be controlled by changing the overlap between the two condensate modes $\varphi_0$ and $\varphi_1$.

The hyperfine states $|0\rangle$ and $|1\rangle$ can be coupled by a two photon transition, where a detuned microwave pulse couples $|0\rangle$ to an intermediate state $|F=2,m_F=0\rangle$, coupled to the final state $|1\rangle$ by a radiofrequency transition \cite{hall}. This two photon transition gives rise to a hopping term $-J(\hat{a}_0^\dagger\hat{a}_1+\hat{a}_1^\dagger\hat{a}_0)$ in the Hamiltonian. In the spin representation this corresponds to the term $-2J\hat{S}_x$, thus with the already known form of the interaction, $\Delta U\hat{S}_z^2$, we recover Hamiltonian \eqref{eq:H}.

To reach optimal parameters, one needs  strong enough interactions and therefore relatively strong confinement. For a  trap frequency $\nu_0=50{\rm Hz}$, and typical scattering length difference  $a_{00}-a_{01}=0.1\, a_{00}=0.53$nm, the interaction strength is $\Delta U/h=0.014$Hz. For atom numbers in the range of $N\sim 2000$, the relevant parameter of the spin model is around $\Delta UN/h=30$Hz. For the typical entropy oscillations plotted in Fig.~\ref{fig:intro2}, the parameter $\alpha=\Delta UN/(2J)$ is roughly $\alpha\sim 2$, corresponding to $J/h=7$Hz. With these parameters, the typical time scale of entropy oscillations and entropy generation for the maximally polarized initial state $|\hat{S}_z=N/2\rangle$ is expected to be around $t\sim 70$ms,which is much shorter than the  lifetime of a condensate, and also much shorter than the coherence time of superposition states \cite{microwave,treutlein,deutsch2,opticaltrap}. Therefore the entropy oscillations should be observable on experimentally realistic time scales.

Let us note that for the parameters above $\overline{U}\gg\Delta U$ implies $N\overline{U}/h\gg\nu_0$. Since the typical scale of interaction energy is much larger than the trapping frequency $\nu_0$, the atoms do not remain in the ground state of the harmonic potential. However, the system can still be described as two coupled single mode condensates with a modified condensate wave function $\varphi_0$, because $N\Delta U/J<\nu_0$ \cite{footnote2}, and the entropy oscillations and entropy generation remain observable with a slightly modified oscillation frequency and entropy production rate.

\section{Conclusions}\label{sec:discuss}

In this work we analyzed the entropy generation for two coupled single-mode Bose-Einstein condensates, realized by loading a condensate into a double well potential, or by an interstate coupling of a two component Bose-Einstein condensate. This system provides one of the simplest examples to study the entropy production by the coherent time evolution of coupled quantum systems. Even though entanglement measurements in generic correlated many-body systems are challenging, in this setting the entanglement between the two potential wells should be experimentally accessible by measuring the number of atoms in the wells. Besides its experimental relevance, the dynamics of coupled single-mode condensates already shows interesting physics. At large particle number imbalances and sufficiently strong interactions, the system enters a self-trapped regime, where the amplitude of population imbalance oscillations gets suppressed due to the large interaction energy of the initial state.

Concentrating on the entropy production during unitary time evolution, we investigated the time dependence of the entropy at $T=0$ temperature, by combining numerical results with analytical calculations. We found that the coherent oscillations of the single mode condensates manifest in entropy oscillations on the top of a steady entropy generation. These coherent oscillations only vanish in the vicinity of the self-trapping transition, where the entropy increases linearly for short time scales. In this pure quantum state, the entropy production originates from the dephasing between different energy eigenstates, eventually leading to a stationary, saturated entropy. Interestingly, this entropy saturation looks like equilibration, in spite of the coherent time evolution of this closed system. These results should be experimentally observable for realistic parameters in microwave measurements with $^{87}$Rb atoms. Here the two modes of the condensate are not spatially separated; instead they correspond to two different hyperfine states of $^{87}$Rb.

 To gain more insight into the entropy oscillations and entropy production on short times scales, we have shown that the time evolution of the entropy can be understood in terms of the semiclassical trajectories of the system. The wave function can be visualized as a broadening packet on the unit sphere, with its center evolving along the classical trajectory. To obtain a quantitative description, we have shown that a Gaussian Ansatz for the wave function, together with a semiclassical expansion, yields a surprisingly good approximation for the exact time evolution.

We also analyzed how the stationary long time limit of the entropy reflects the semiclassical dynamics of the system. The classical fixed point is revealed by a local minimum in the entropy, related to the strong confinement of trajectories in the vicinity of this point, while the self-trapping transition is accompanied by a sudden entropy jump of size $\log 2$, due to the rapid change by a factor of 2 in the length of trajectories.

In order to investigate the dephasing induced equilibration of the entropy in more detail, we compared the numerical results to the prediction of a classical microcanonical ensemble, where the spin vector is distributed uniformly over the classical trajectory. We found that this ensemble yields a surprisingly accurate description for the stationary limit of the entropy. To gain more insight into the exact time evolution of the system, we calculated the overlap of the wave function with the spin coherent states of different orientations. We have shown that this overlap traces out the classical trajectories on the unit sphere at long times, supporting our picture describing the system in terms of a classical microcanonical ensemble. 

In this work we concentrated on the entropy generation in a pure state, and performed all calculations at $T=0$ temperature. Thermal fluctuations are expected to shift the entropy of the system to higher values, while the reduction of coherence starts to wash out the oscillations during the entropy production. However, the entropy oscillations should still remain visible for low enough temperatures of the order of a few hundred nK. 

Since entropy generation lies at the heart of equilibration and thermalization in closed systems, the detailed analysis of entropy production in other correlated many-body systems, and entanglement spreading in the presence of conserved quantities, remains a question of fundamental interest. 
%{\bf Other systems??}

\begin{acknowledgments}
We acknowledge fruitful discussions with P\'eter Domokos. This research has been  supported by the National Research, Development and Innovation Office - NKFIH Nos. K105149 and SNN118028. ED acknowledges support from Harvard-MIT CUA, NSF Grant No. DMR-1308435, AFOSR Quantum Simulation MURI, AFOSR Grant No. FA9550-16-1-0323.  
\end{acknowledgments}

\appendix

\section{Classical microcanonical distribution of $\Omega_z$}\label{app:PSz}

In this appendix we outline the calculation of the continuous distribution $P(\Omega_z)$ used in Eq. \eqref{eq:Sclass}. In a microcanonical description, we assume that the unit vector $\vec{\Omega}$ is distributed uniformly along the curve of constant energy, selected by the initial state $\vec{\Omega}=(\sin\theta,\,0,\,\cos\theta)$. The projection of this classical trajectory to the $x-y$ plane is a circle, given by 
\begin{equation*}
\left(\Omega_x+\dfrac{1}{\alpha}\right)^2+\Omega_y^2=r_\theta^2,
\end{equation*}
with $r_\theta=|\sin\theta+1/\alpha|$. Based on this expression, the trajectory can be parametrized by an angle $\chi$  as
\begin{align*}
&\vec{\Omega}(\chi)=\\
&\left(r_\theta\cos\chi-\frac{1}{\alpha},\,r_\theta\sin\chi,\sqrt{1-r_\theta^2-\frac{1}{\alpha^2}+\frac{2 r_\theta}{\alpha}\cos\chi}\right).
\end{align*}
The length of a small arc segment, $\Delta s$, can be expressed with the parameter change $\Delta\chi$ as
\begin{equation*}
\Delta s=|\partial_\chi\vec{\Omega}(\chi)|\Delta\chi,
\end{equation*}
with $|\partial_\chi\vec{\Omega}(\chi)|$ denoting the length of the tangent vector $\partial_\chi\vec{\Omega}(\chi)$. For a uniform distribution along the curve, $\vec{\Omega}$ points to this segment with probability $\Delta s/s_0$, where $s_0$ is the total length of the trajectory. This results in the following distribution $P(\chi)$ for parameter $\chi$
\begin{equation*}
P(\chi)\Delta\chi=\frac{\Delta s}{s_0}\;\Rightarrow\; P(\chi)=\frac{1}{s_0}|\partial_\chi\vec{\Omega}(\chi)|.
\end{equation*}
The distribution of $\Omega_z$ can be expressed as
\begin{equation*}
P(\Omega_z)=\left. 2 P(\chi)\left(\frac{\partial\Omega_z}{\partial\chi}\right)^{-1}\right|_{\chi(\Omega_z)},
\end{equation*}
where $\chi(\Omega_z)$ denotes the inverse function of $\Omega_z(\chi)$, and the factor 2 arises from the symmetry $\Omega_z(\chi)=\Omega_z(-\chi)$.

The distribution $P(\Omega_z)$ as a function of parameter $\alpha$ and initial condition $\theta$ is given by
\begin{align*}
&P(\Omega_z)=\\
&\frac{2}{s_0}\sqrt{1+\dfrac{\alpha^2\Omega_z^2}{1-\left[\dfrac{\alpha}{2 r_\theta}(\Omega_z^2-\cos^2\theta)+{\rm sgn}(\sin\theta+1/\alpha)\right]^2}},
\end{align*}
with sgn denoting the sign function. The length of the trajectory, $s_0$, ensures the correct normalization of the distribution,
\begin{equation*}
\int_{\Omega_z^{\rm min}}^{\Omega_z^{\rm max}}{\rm d}\Omega_z\, P(\Omega_z)=1.
\end{equation*}
 The support of this distribution $(\Omega_z^{\rm min},\,\Omega_z^{\rm max})$, however, is different on the two sides of the self-trapping transition. In the non-trapped regime 
$$(\Omega_z^{\rm min},\,\Omega_z^{\rm max})=(-\cos\theta,\,\cos\theta),$$
while for a self-trapped state
\begin{align*}
&\Omega_z^{\rm min}=\sqrt{\cos^2\theta-\dfrac{2}{\alpha}(r_\theta+\sin\theta+1/\alpha)},\\
&\Omega_z^{\rm min}=\sqrt{\cos^2\theta+\dfrac{2}{\alpha}(r_\theta+\sin\theta+1/\alpha)}.
\end{align*}


\begin{references}
\bibitem{footnote0}Let us note that certain entanglement measures, characterizing that portion of the von Neumann entropy which can be used as a resource in quantum information theory, are always zero for a two-mode system with fixed total particle number \cite{QIresource}. Instead of quantifying quantum resource, in our case the von Neumann entropy given by Eq. \eqref{eq:Sn} measures the coherent, entangled fluctuations of the particle numbers in the two potential wells.
\bibitem{footnote}Let us note that interpreting Eq. \eqref{eq:S} as an entanglement entropy is no longer obvious in the spin representation. Instead, $\mathcal{S}(t)$ is an entropy associated with the observable $\hat{S}_z$, defined through the full distribution $P_t(m)$.
\bibitem{footnote2}For strong interactions $\alpha\gg 1$, the splitting of the doublet with the highest energy is given by $\Delta E_{\rm doublet}/U\approx 2N(J/U)^N/(N-1)!$ \cite{tunneling}.
\bibitem{footnote4}Note that the sharp quantum interference minima, appearing on the unit sphere in Fig.~\ref{fig:overlap}, are absent from Fig.~\ref{fig:tunnel}, because in Fig.~\ref{fig:tunnel} we plot the state of the system at two special times. Both the initial state of the system and the state after very long time, when almost all particles have tunneled to the other potential well, are well described by a coherent state, thus the sharp minima on the unit sphere are washed out.
\end{references}
\end{document}